\documentclass[lettersize,journal]{IEEEtran}
\usepackage{amsmath,amsfonts}
\usepackage{algorithmic}
\usepackage{array}
\usepackage[caption=false,font=normalsize,labelfont=sf,textfont=sf]{subfig}
\usepackage{textcomp}
\usepackage{stfloats}
\usepackage{url}
\usepackage{verbatim}
\usepackage{graphicx}

\usepackage{booktabs}
\usepackage{multirow}
\usepackage{color}

\def\BibTeX{{\rm B\kern-.05em{\sc i\kern-.025em b}\kern-.08em
    T\kern-.1667em\lower.7ex\hbox{E}\kern-.125emX}}
\usepackage{balance}
\begin{document}

\title{Rethinking Vision Transformer for Large-Scale Fine-Grained Image Retrieval}

\author{Xin~Jiang, 
        Hao~Tang, \emph{Member, IEEE}, 
        Yonghua~Pan, 
        and Zechao~Li, \emph{Senior Member, IEEE}% <-this % stops a space
\thanks{This work was supported by National Natural Science Foundation of China (Grant No. 62425603), and Basic Research Program of Jiangsu Province (Grant No. BK20240011).}
\thanks{X. Jiang and Z. Li are with the School of Computer Science and Engineering, Nanjing University of Science and Technology, Nanjing 210094, China  (e-mail: xinjiang@njust.edu.cn; zechao.li@njust.edu.cn). \emph{(Corresponding Author: Zechao Li.)}}
\thanks{H. Tang is with the Centre for Smart Health, The Hong Kong Polytechnic University, Hong Kong, China (e-mail:~howard.haotang@gmail.com). Y. Pan is with Guangxi Academy of Sciences, Nanning, Guangxi, China (e-mail:~yhpan@gxas.cn).}
}

\markboth{Journal of \LaTeX\ Class Files,~Vol.~18, No.~9, September~2020}%
{How to Use the IEEEtran \LaTeX \ Templates}

\maketitle

\begin{abstract}
Large-scale fine-grained image retrieval (FGIR) aims to  retrieve images belonging to the same subcategory as a given query by capturing subtle differences in a large-scale setting. Recently, Vision Transformers (ViT) have been employed in FGIR due to their powerful self-attention mechanism for modeling long-range dependencies. However, most Transformer-based methods focus primarily on leveraging self-attention to distinguish fine-grained details, while overlooking the high computational complexity and redundant dependencies inherent to these models, limiting their scalability and effectiveness in large-scale FGIR. 
In this paper, we propose an Efficient and Effective ViT-based framework, termed \textbf{EET}, which integrates token pruning module with a discriminative transfer strategy to address these limitations. Specifically, we introduce a content-based token pruning scheme to enhance the efficiency of the vanilla ViT, progressively removing background or low-discriminative tokens at different stages by exploiting feature responses and self-attention mechanism. To ensure the resulting efficient ViT retains strong discriminative power, we further present a discriminative transfer strategy comprising both \textit{discriminative knowledge transfer} and \textit{discriminative region guidance}. Using a distillation paradigm, these components transfer knowledge from a larger ``teacher'' ViT to a more efficient ``student'' model, guiding the latter to focus on subtle yet crucial regions in a cost-free manner. 
Extensive experiments on two widely-used fine-grained datasets and four large-scale fine-grained datasets demonstrate the effectiveness of our method. Specifically, EET reduces the inference latency of ViT-Small by 42.7\% and boosts the retrieval performance of 16-bit hash codes by 5.15\% on the challenging NABirds dataset. The code is publicly available at: \url{https://github.com/WhiteJiang/EET}.
\end{abstract}


\section{Introduction}
\label{s1}
\begin{figure}
    \centering
    \includegraphics[width=0.95\linewidth]{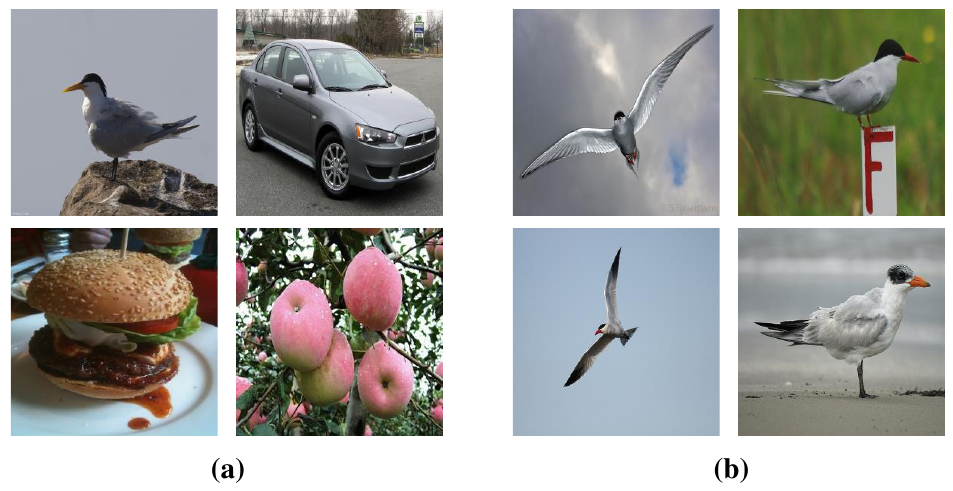}
    \caption{(a) Coarse-grained images: Significant visual differences between images from different categories. (b) Fine-grained images: Large intra-class variations within each row, and small inter-class variations within each column. This unique characteristic poses challenges when transitioning from coarse-grained to fine-grained hashing retrieval.}
    \label{fig:fig0}
\end{figure}

\IEEEPARstart {F}{ine-grained} image retrieval~(FGIR) is a fundamental task in computer vision~\cite{WeiSAWPTYB22} and multimedia~\cite{LiTM19}. Its goal is to retrieve images belonging to the same subcategory within a dataset containing multiple subcategories under a broader meta-category (\emph{e.g.,} cars~\cite{car}, birds~\cite{wah2011caltech}).
Compared with coarse-grained image retrieval, FGIR poses greater challenges due to small inter-class variations and large intra-class variations, as shown in Figure~\ref{fig:fig0}. Furthermore, FGIR must rank all instances based on subtle visual details in the query, placing the most relevant images at the top. However, the rapid growth of fine-grained data on the internet has made traditional methods, which rely on high-dimensional features~\cite{dvf,hist,proxy}, computationally expensive and difficult to scale. To address this, hashing-based methods~\cite{oneloss,dahnet,dpn,csq} have attracted increasing attention by converting high-dimensional features into compact binary codes, thus reducing both computation and storage overhead. Most existing hashing frameworks, however, are designed for coarse-grained images (Figure~\ref{fig:fig0}(a)), where overall visual differences are prominent. Such methods often fail in the fine-grained setting (Figure~\ref{fig:fig0}(b)), where images appear highly similar, making it difficult to capture subtle yet crucial distinctions.

Recent fine-grained hashing methods~\cite{dahnet,semicon,anet++} have made significant progress by incorporating complicated attention mechanisms to capture these subtle differences. Most rely on convolutional neural networks (CNNs), which can be limited in representing fine-grained patterns for large-scale retrieval. Moreover, many state-of-the-art (SOTA) methods cannot balance effectiveness and efficiency. For example, \textsc{DAHNet}~\cite{dahnet} leverages attention mechanisms to establish association between global/local image features and  hash bits, yet it nearly doubles inference time compared to its baseline, thus compromising efficiency in large-scale FGIR scenarios. Meanwhile, Vision Transformers (ViTs)~\cite{vit,deit} have demonstrate powerful performance in various computer vision tasks by modeling patch-wise dependencies with multi-head self-attention (MHSA). Their global receptive field is advantageous for capturing fine-grained details. However, MHSA incurs quadratic computational complexity with respect to the number of image tokens, resulting in higher latency than CNNs, which is an issue that becomes critical in large-scale FGIR. Existing ViT-based FGIR methods~\cite{dvf,msvit,swinfghash} often ignore this efficiency bottleneck, sometimes adding extra computational burdens to achieve more discriminative representations. Thus, designing an efficient yet effective ViT-based framework for large-scale FGIR remains an open challenge.

The human visual perception system adopts a top-down cognitive mechanism~\cite{maier2019no,TangYLT22,tang2024divide} to identify objects through a global-to-local process. As objects become harder to distinguish from similar categories, the visual system focuses on subtle, discriminative regions while ignoring background areas or low-discriminative regions. Inspired by this, we aim for ViT to emulate this top-down cognitive process by gradually reducing unnecessary image token computations during inference, thereby making ViT more suitable for large-scale FGIR tasks. Recent studies~\cite{ats,BolyaFDZFH23,DynamicViT} have explored token pruning as a means to reduce computation, but these methods are primarily designed for coarse-grained images and often sacrifice performance. The question, then, is how to prune tokens to retain discriminative patches crucial for fine-grained tasks, without compromising retrieval performance in large-scale scenarios.

In this paper, we propose an \textbf{efficient and effective ViT-based framework} called \textbf{EET} for large-scale FGIR. As shown in Figure~\ref{fig:overview}, our framework integrates two main components: (1) \textbf{Content-based Token Pruning (CTP) scheme:} CTP leverages the importance of tokens within the MHSA intermediate token content to adjust the class attention scores, identifying tokens that contain subtle yet discriminative differences. It progressively removes background and low-discriminative tokens in a hierarchical manner, mimicking the human global-to-local attention process and substantially reducing computational overhead. (2) \textbf{Discriminative Transfer Strategy:} While pruning improves inference efficiency, it can degrade the fine-grained discriminative power of the pruned ViT. To address this issue, we introduce \textit{Discriminative Knowledge Transfer (DKT)} and \textit{Discriminative Region Guidance (DRG)}. Here, DKT employs a heavier vanilla  ``teacher'' ViT to distill rich visual knowledge into the more efficient pruned ``student'' ViT, while DRG further guides the student to focus on subtle yet critical regions, thereby preserving strong discriminative capacity despite token reduction.
Extensive experiments on two widely used fine-grained datasets and four large-scale fine-grained datasets demonstrate the effectiveness of our EET framework. In particular, EET reduces inference latency by over $42\%$ on ViT-Small while maintaining strong discriminative capability for retrieval.

In summary, the main contributions of this paper are as follows:
\begin{itemize}
    \item We investigate the challenges of deploying ViT for large-scale fine-grained image retrieval and propose EET, an efficient and effective ViT-based framework suitable for deployment.
    \item We propose a content-based token pruning scheme combined with a discriminative transfer strategy to enable ViT to efficiently and effectively capture fine-grained differences between similar objects.
    \item Experimental results on six fine-grained image retrieval benchmarks demonstrate the superior performance of EET. Specifically, EET reduces inference latency by over 42\% for ViT-Small.
\end{itemize}

The rest of the paper is organized as follows: Section~\ref{s2} reviews related work; Section~\ref{s3} provides a preliminary introduction to vision transformers; Section~\ref{method} details the proposed method; Section~\ref{s5} presents experimental results and analysis; finally, Section~\ref{s6} concludes the paper.
\section{Related Works} \label{s2}
\subsection{Fine-grained Image Classification}
Fine-grained image classification, as an upstream task of fine-grained image retrieval, has made significant progress in addressing fine-grained challenges. Current methods primarily focus on two research directions: feature encoding~\cite{LinRM15,GaoBZD16,YuZZZY18} and part localization~\cite{LinSLJ15,zha2023boosting, IELT,mpfgvc}.

Feature encoding methods aim to enhance feature learning by combining distinct features. More specifically, B-CNN~\cite{LinRM15} uses two feature extractors to extract features from a single image and computes the outer product of corresponding points to derive the final feature representation. Compact B-CNN~\cite{GaoBZD16} introduces compact bilinear pooling, which effectively reduces feature dimensions while maintaining high performance. HBP~\cite{YuZZZY18} introduces a hierarchical bilinear pooling framework that leverages supplementary information from intermediate convolutional layers. This method improves performance by integrating multiple cross-layer bilinear modules. Although feature encoding methods enhance the generalization performance of network models in fine-grained image classification, they often overlook subtle yet semantically rich regions in fine-grained images.
In contrast to feature encoding methods, part localization methods focus on identifying discriminative regions to distinguish subtle inter-class differences. Specifically, part localization methods can be classified into two types. The first type, as described in R-CNN~\cite{ZhangDGD14} and LAC~\cite{LinSLJ15}, utilizes bounding box annotations to detect discriminative regions and generate discriminative feature representations.
However, obtaining bounding box annotations can incur high costs. Consequently, the second category of methods~\cite{HanYCFX22, LiuZBZZ22,mpfgvc,shen2024imagdressing} utilizes weakly supervised techniques to identify discriminative regions.

\subsection{Fine-grained Image Retrieval}
Fine-grained image retrieval~(FGIR), an integral component of fine-grained image analysis~\cite{WeiSAWPTYB22}, has garnered increasing attention in recent years. It aims to distinguish between visually similar sub-categories by identifying subtle differences in their details. Unlike coarse-grained image retrieval, the main challenge in FGIR is the small inter-class variations and large intra-class variations observed in both database and query images. To tackle this challenge, existing FGIR approaches can be broadly categorized into two groups. The first group, encoding-based schemes, aims to learn an embedding space where samples from the same subcategory are drawn closer while those from different subcategories are pushed apart~\cite{hyp,proxy,proxynca}. The second group of methods, referred to as location-based schemes, focuses on training a subnetwork to identify discriminative regions or devising effective strategies for extracting relevant object features to enhance the retrieval process~\cite{dvf,song2023boosting,gao2024re,git}.
For example, DVF~\cite{dvf} introduces a visual foundation model to enable zero-shot object localization and semantic token filter module to locate discriminative tokens, effectively reducing the impact of background noise. DToP~\cite{song2023boosting} designs a local branch to identify important patch tokens and enhance their significance. However, both methods introduce significant computational overhead and struggle with large-scale data processing.

In addition, they face substantial storage costs when handling large-scale data. To alleviate this issue, fine-grained hashing, a technique that creates concise binary codes to represent fine-grained images, has recently garnered significant attention in the fine-grained community~\cite{dsah,anet,fish,dbaq}. Specifically, DSaH~\cite{dsah} emerges as the pioneering method designed for the fine-grained hashing problem. It utilizes an attention mechanism to automatically identify discriminative regions and extract distinguishing features for generating concise hash codes. FISH~\cite{fish} introduces a double-filtering mechanism for fine-grained feature extraction and refinement, along with a proxy-based loss function to capture class-level characteristics.
MSViT~\cite{msvit} introduces a dual-branch vision transformer to process image patches with different granularities, thereby perceiving local features to enhance the discriminability of the model.
However, their focus on designing effective yet inefficient modules to enhance performance hampers their efficiency in large-scale FGIR. In response, we propose an efficient and effective vision transformer framework for solving large-scale FGIR tasks.

\subsection{Learning to Hash}
Learning to hash, a fundamental component of approximate nearest neighbor search that transforms data items into short binary codes, has emerged as a promising approach for addressing large-scale image retrieval tasks~\cite{QinXZWH24},\emph{ e.g.,} face image retrieval~\cite{c4}, social image retrieval~\cite{c5}, etc. Research in hashing can be categorized into two groups: data-independent hashing~\cite{c8,XuCXLZYY23} and data-dependent hashing~\cite{itq,adsh,LuJLT24,CFBH}. Specifically, the data-independent hashing methods aim to refine hash learning from various perspectives. For instance, they have proposed random hash functions that satisfy the local sensitive property~\cite{c8}, improved search schemes~\cite{c7}, and enhanced the computational efficiency of hash functions~\cite{c8}, among others. In contrast with data-independent hashing methods, data-dependent hashing methods leverage advancements in deep learning, integrating hashing learning into an end-to-end framework based on deep networks to preserve similarity~\cite{itq,adsh,LuJLT24}. Considering the complex visual characteristics in fine-grained scenarios, we investigate the efficacy of data-dependent hashing for large-scale FGIR by integrating feature learning and hash code learning into a unified end-to-end framework.

\subsection{Vision Transformer Pruning}
While ViT delivers remarkable results in comparison to CNN in the computer vision field, it also entails higher computational costs. To enhance the efficiency of ViT, existing approaches can be separated into two groups: static ViT pruning~\cite{c11,c10} and dynamic ViT pruning~\cite{c14,c15,c16}. The former focuses on parameter compression. For instance, NViT~\cite{c11} identifies the importance of global weights through Taylor expansion on the loss function, followed by structured pruning and parameter reassignment based on dimensional trends. SViTE~\cite{c10}, on the other hand, extensively exploits the sparsity of ViT by employing structured pruning, unstructured sparsity, and token pruning. The latter benefits from the Transformer's parallel computing mechanism to accelerate inference by pruning image tokens. DynamicViT~\cite{DynamicViT} and IA-RED$^2$~\cite{c14} score tokens and discard unimportant ones by integrating prediction modules. EViT~\cite{c15} and Evo-ViT~\cite{c16} utilize the class attention score to assess the informativeness of each token and discard those deemed unimportant. However, existing methods mainly target coarse-grained images and focus more on efficiency than effectiveness. To address this, we introduce a discriminative transfer strategy that incurs no inference burden, combined with token pruning to preserve tokens containing subtle yet discriminative differences, thereby improving both efficiency and effectiveness.

\section{Preliminary of Vision Transformer} \label{s3}
Given an input image $\mathbf{X}$, the Vision Transformer (ViT)~\cite{vit} initially partitions the image into $N = N_{h} \times N_{w}$ non-overlapping patches, each of size $P \times P$. Here, $N_{h}$ and $N_{w}$ are the numbers of patches along the image’s height and width, respectively. Subsequently, these patches are projected into embedding tokens $\mathbf{E} = [\mathbf{E}^{1}, \mathbf{E}^{2}, \dots, \mathbf{E}^{N}] \in \mathbb{R}^{N \times D}$ through a learnable
linear projection $\mathbf{P}_{emb} \in \mathbb{R}^{P^{2}\times D}$, where $D$ denotes the embedding dimension per token. A special class token $\mathbf{E}_{class} \in \mathbb{R}^{D}$ is then prepended and a position embedding $\mathbf{E}_{pos} \in \mathbb{R}^{(N+1) \times D}$ is added element-wise to form the initial input token sequence as $\mathbf{E}_{0} = 	\lbrack \mathbf{E}_{class}, \mathbf{E}^{1}, \mathbf{E}^{2}, \cdots, \mathbf{E}^{N} \rbrack + \mathbf{E}_{pos}$. 

The ViT encoder comprises $L$ transformer layers, each containing a multi-head self-attention (MHSA) module and a multi-layer perception (MLP). For the $i$-th layer, given an input token sequence $\mathbf{E}_{i-1}$, the output $\mathbf{E}_{i}$ is computed as follows:
\begin{equation}
\begin{aligned}
 \mathbf{E}_{i}^{\prime} &= \mathbf{E}_{i-1} + \textrm{MHSA}(\mathrm{LN}(\mathbf{E}_{i-1})), \\
    \mathbf{E}_{i} &= \mathbf{E}_{i}^{\prime} + \textrm{MLP}(\mathrm{LN}(\mathbf{E}_{i}^{\prime})),
\end{aligned}
\end{equation}
where $i\in \lbrace1, 2, \cdots, L\rbrace$, and $\mathrm{LN}(\cdot)$ denotes layer normalization.
The \textrm{MHSA} module projects $\mathbf{E}_{i-1}$ into queries $\mathbf{Q}$, keys $\mathbf{K}$, and values $\mathbf{V}$, each in $\mathbb{R}^{(N+1)\times D}$. These projections are then split into $H$ heads to enable parallel attention:
\begin{equation}
   \mathrm{Attention}(\mathbf{Q}, \mathbf{K}, \mathbf{V}) = \mathrm{softmax}\Bigl(\frac{\mathbf{Q}\mathbf{K}^\top}{\sqrt{D_{h}}}\Bigr)\mathbf{V},
   \label{mhsa}
\end{equation} 
where $D_{h} = \frac{D}{H}$.
Notably, the attention scores between the class token $\mathbf{E}_{class}$ and the remaining tokens can be extracted to indicate which patches contribute most to classification as:
\begin{equation}
    \mathrm{A}\bigl(\mathbf{E}_{class},:\bigr) = \mathrm{softmax}\Bigl(\frac{\mathbf{Q}_{[0]}\mathbf{K}_{[1:]}}{\sqrt{D_{h}}}\Bigr) \in \mathbb{R}^{H \times N}.
    \label{mhsa_v1}
\end{equation}
\section{Methods} \label{method}
\begin{figure*}
    \centering
    \includegraphics[width=0.95\textwidth]{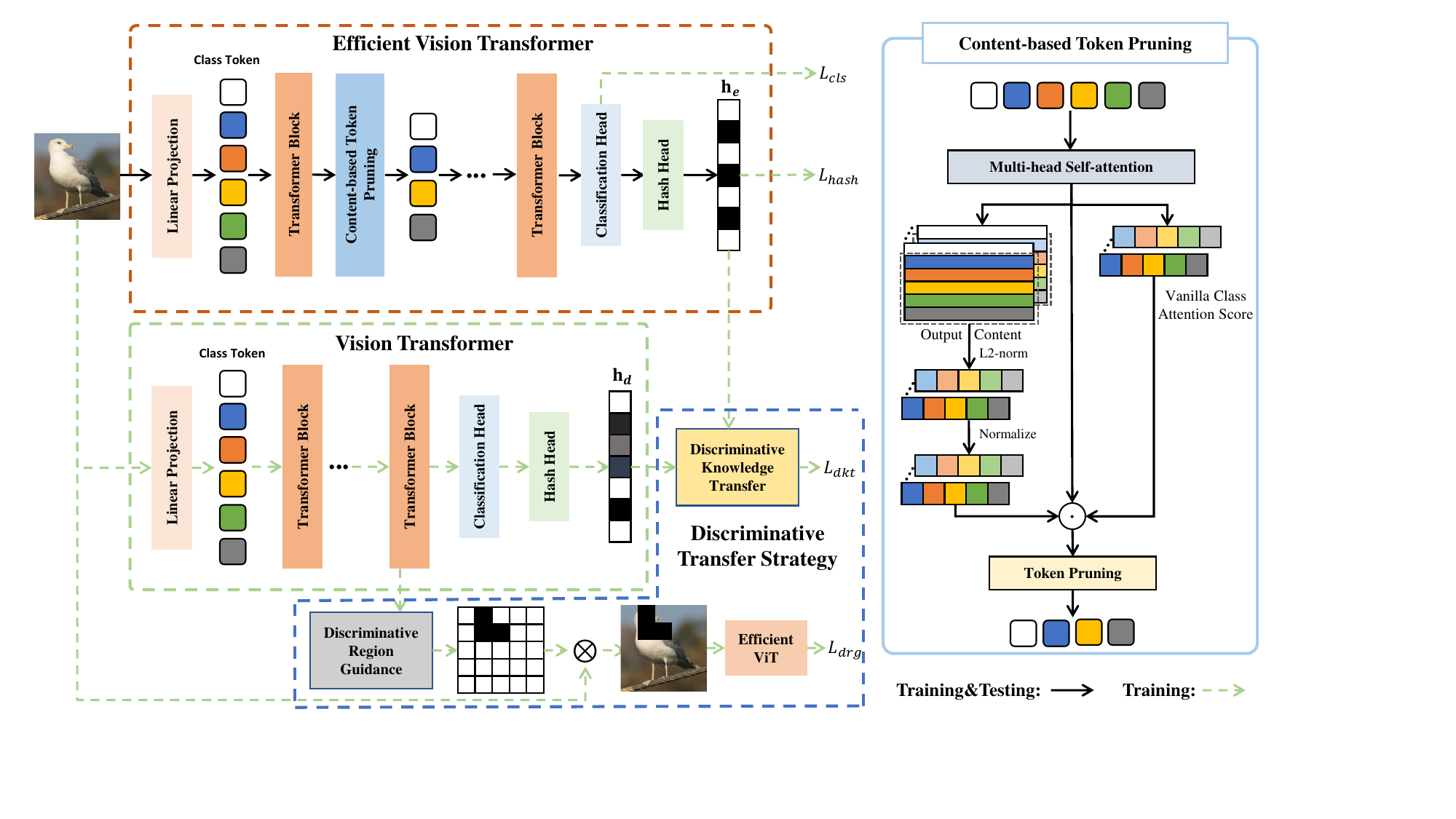}
    \caption{Overview of the proposed framework, which comprises three core components: \emph{\textbf{(1) Content-based Token Pruning (CTP)}}, 
    \emph{\textbf{(2) Discriminative Knowledge Transfer (DKT)}},
    and 
    \emph{\textbf{(3) Discriminative Region Guidance (DRG)}}.
    CTP progressively discards background and low-discriminative tokens to significantly improve the computational efficiency of the Vision Transformer (ViT). 
    The discriminative transfer strategy, consisting of DKT and DRG, enables the efficient ViT to learn highly discriminative hash code representations in a cost-free way. 
    During inference, only the efficient ViT (\emph{i.e.,}~with pruned tokens) is employed for hash code generation, thereby maintaining high efficiency.}
    
    \label{fig:overview}
\end{figure*}
\subsection{Overall Framework and Notations}
The overall structure of EET is illustrated in Figure~\ref{fig:overview}. 
It consists of two models, \emph{i.e.,} \textit{Efficient Vision Transformer~(EViT)} and the standard \textit{Vision Transformer~(ViT)}, and three key modules: \emph{\textbf{(1) Content-based Token Pruning (CTP)}}, \emph{\textbf{(2) Discriminative Knowledge Transfer (DKT)}}, and \emph{\textbf{(3) Discriminative Region Guidance (DRG)}}. 
By hierarchically integrating CTP into ViT, we form the EViT, which enables efficient processing of large-scale, fine-grained data. Meanwhile, DKT and DRG are employed to enhance EViT’s discriminative ability for fine-grained objects in a cost-free manner. During inference, only the EViT is utilized to generate hash codes.

Formally, let $\mathbf{X}$ be an input image and $\mathbf{T(\cdot)}$ denote a backbone network that encodes $\mathbf{X}$ into a $D$-dimensional embedding $\mathbf{E}_{class} \in \mathbb{R}^{D}$. Thus,
\begin{equation} \mathbf{E}_{\mathrm{class}} = \mathbf{T}(\mathbf{X}). 
\end{equation}
For efficient retrieval, the final feature $\mathbf{E}_{class}$
is projected into a $k$-bit hash code $\mathbf{b} \in \lbrace -1, 1 \rbrace^{k}$ by applying a hash projection operation followed by the element-wise sign function $\mathrm{sign}(\cdot)$:
\begin{equation}
    \mathrm{sign}(x) = \begin{cases}
-1, & \quad x \leq 0, \\
1, & \quad x > 0.
\end{cases}
\end{equation}
This process yields binary codes that facilitate highly efficient similarity search in large-scale databases.

\subsection{Content-based Token Pruning} ~\label{mctc}

The efficiency of a Vision Transformer (ViT) is inversely proportional to the number of tokens it processes. Previous works~\cite{mpfgvc,eViT,c14} on coarse-grained classification leverage token-pruning strategies to reduce token count and thus improve ViT efficiency. However, in the context of fine-grained images, pruning must not only discard redundant tokens but also preserve subtle yet highly discriminative details. Furthermore, existing pruning approaches often average attention scores across multiple heads~\cite{mpfgvc,eViT,c14}, overlooking the distinct information each head can capture. Since the class token mostly attends to the most salient regions, it may miss subtle differences critical for fine-grained recognition. 
To address these challenges, we introduce a \emph{Content-based Token Pruning} (CTP) scheme. Rather than relying solely on averaged attention scores, CTP leverages the intermediate token content from the multi-head self-attention (MHSA) module to assign different weights across attention heads. This approach effectively identifies subtle but discriminative tokens without adding extra computational cost or parameters.

\subsubsection{Intermediate Token Content}
Let $\textrm{Content}^{h,l}_{i} \in \mathbb{R}^{D_{h}}$ denote the intermediate token content produced by the $h$-th head in the $l$-th layer of MHSA (see Eq.(\ref{mhsa})). Inspired by SENet~\cite{senet}, which uses feature-response magnitude to gauge feature importance, we compute the importance of the $i$-th token as:
\begin{equation}
    S^{h,l}_{i} = \lVert\textrm{Content}^{h,l}_{i}\rVert_{2},
\end{equation}
where $\lVert\cdot\rVert_{2}$ denotes the $l_{2}$-norm.
Next, we normalize $S^{h,l}_{i}$ across all attention heads to obtain a weighting factor:
\begin{equation}
    W^{h,l}_{i} = \frac{S^{h,l}_{i}}{\sum_{h=1}^H S^{h,l}_{i}}.
\end{equation}
This factor $W^{h,l}$, in conjunction with the attention score $\mathrm{A}^{h,l}$ (see Eq.(\ref{mhsa_v1})) between the class token and other tokens, pinpoints both salient and subtle discriminative tokens:
\begin{equation} \label{ti}
    M^{l} = \sum_{h=1}^H W^{h,l} \cdot \mathrm{A}^{h,l}.
\end{equation}

\subsubsection{Hierarchical Token Pruning}
Finally, we retain the top $N_\omega \cdot \mathrm{len}(M^{l})$ tokens in the $l$-th layer, as determined by $M^{l}$, where $N_\omega$ is a hyper-parameter controlling the pruning rate. In practice, we insert CTP after the $4$-th, $8$-th, and $10$-th transformer layers in ViT, mimicking a human-like top-down recognition process by progressively discarding unimportant tokens. During inference, only the pruned subset of tokens flows into subsequent layers, thereby improving computational efficiency without sacrificing the fine-grained details crucial for recognition.

\subsection{Discriminative Transfer Strategy}
The efficiency of ViT can be significantly increased via content-based token pruning. However, this pruning may also reduce the model’s ability to detect subtle differences in fine-grained images. To address this limitation, we propose a \emph{discriminative transfer strategy}, composed of \emph{Discriminative Knowledge Transfer} (DKT) and \emph{Discriminative Region Guidance} (DRG), to train and optimize the EViT. This strategy enriches EViT’s discriminative power for fine-grained subcategories without adding computational overhead during inference, since both DKT and DRG are only employed during training.

\subsubsection{Discriminative Knowledge Transfer}  \label{dkt}
The hash codes $\mathbf{h}_e$ produced by the EViT are derived from a pruned (\emph{i.e.,}~incomplete) set of tokens, potentially causing a loss of fine-grained discriminative information. Recent studies~\cite{vitkd,lv2022transformer} have shown that knowledge distillation~\cite{distill} can mitigate such information loss. However, these methods primarily target coarse-grained classification. For instance,  ViTKD~\cite{vitkd} requires symmetric ViT architectures, whereas our EET is asymmetric. Meanwhile, the approach in~\cite{lv2022transformer} aligns hash code distributions within the same category but may inadvertently narrow the overall distribution, making it more difficult to correct misassigned hash codes for individual fine-grained objects. To address this issues, we propose \emph{Discriminative Knowledge Transfer} (DKT), which extends knowledge distillation to a retrieval setting by mapping images into a hash-code space. Specifically, we minimize the cosine distance between hash codes produced by the standard ViT ($\mathbf{h}_d$) and the EViT ($\mathbf{h}_e$).

In hash-based retrieval, one typically uses the Hamming distance between binary codes $\mathbf{b}_i$ and $\mathbf{b}_j$. Converting continuous-valued hash codes $\mathbf{h}_i$ and $\mathbf{h}_j$ into binary codes $\mathbf{b}_i$ and $\mathbf{b}_j$ involves non-differentiable sign operations. However, the Hamming distance between binary codes can be interpreted as the cosine distance~\cite{oneloss}. Specifically, the cosine similarity between hash codes $\mathbf{h}_i$ and $\mathbf{h}_j$ can approximate the Hamming distance between binary codes $\mathbf{b}_i$ and $\mathbf{b}_j$ as follows:
\begin{equation}
    \mathrm{hamm}(\mathbf{b}_i, \mathbf{b}_j)\simeq \frac{k}{2}\bigl(1 - \cos(\mathbf{h}_i, \mathbf{h}_j)\bigr),
\end{equation}
where $\mathrm{hamm}(\cdot)$ denotes Hamming distance function, $k$ is the hash code length, and $\mathbf{b}_i = \mathrm{sign}(\mathbf{h}_i)$. 
Consequently, to transfer discriminative visual knowledge from ViT into EViT, we define the DKT loss as:
\begin{equation}
    \mathcal{L}_{\mathrm{dkt}} = 1 - \cos(\mathbf{h}_e, \mathbf{h}_d),
\end{equation}
where $\mathbf{h}_d$ and $\mathbf{h}_e$ are the ViT and EViT hash codes, respectively. By minimizing $\mathcal{L}_{dkt}$, the EViT acquires the discriminative capability of the standard ViT, enhancing retrieval performance for visually similar objects.

\subsubsection{Discriminative Region Guidance} ~\label{drg}
As mentioned in Section~\ref{s1}, the the FGIR task must handle small inter-class and large intra-class variations. Therefore, simply identifying the most discriminative regions is insufficient. Consequently, we introduce \emph{Discriminative Region Guidance} (DRG), which draws on the concept of ``masking'' to remove the most salient regions in the original image, forcing the EViT to locate more subtle discriminative cues. It is worth noting that, unlike more complex, high-capacity models~\cite{dahnet,semicon}, DRG is lightweight and adds no extra inference costs.

Let $\mathbf{M}^{L}_{t} \in \mathbb{R}^{N}$ be the token-importance map from the $L$-th layer of the ViT. We create a binary mask $\hat{\mathbf{M}}$ by setting the top $K$ salient locations to $0$ and the rest to $1$:
\begin{equation}
\hat{\mathbf{M}}_{i} = 
\begin{cases}
	0, & \quad \text{if}~i \in \mathrm{TopK}(\mathbf{M}^{L}_{t}), \\
	1, & \quad \text{otherwise}. 
\end{cases}
\end{equation}
where $\textrm{TopK}(\cdot)$ is a function that returns the indices of the top $K$ salient regions. We then resize $\hat{\mathbf{M}}$ to match the spatial dimensions of the input image by replicating each element $P \times P$ times. The resulting masked image $\mathbf{X}'$ is given by $ \mathbf{X}' = \mathbf{X} \otimes \hat{\mathbf{M}}$, which mask the most discriminative region. Finally, we apply a standard classification loss to guide the EViT to attend to finer details:
\begin{equation}
    \mathcal{L}_{\mathrm{drg}} = \mathrm{CE}\bigl(\hat{y}', y\bigr),
\end{equation}
where $\mathrm{CE}(\cdot)$ denotes the cross-entropy loss, and $y$ and $\hat{y}'$ are the ground-truth label and the classification prediction of the masked image $\mathbf{X}'$, respectively. By masking out salient regions, the EViT is compelled to exploit additional subtle cues for improved fine-grained recognition.

\subsection{Hash Code Learning}
As mentioned earlier, the core idea of this paper is to design an efficient and effective hashing framework for large-scale FGIR, so the hash loss function design is not our focus. Therefore, we implement proxy-based hash code learning to ensure efficient retrieval and training as proposed in FISH~\cite{fish}.
Specifically, the optimization process of hash code is divided into two steps.
The learning process of the hash code is divided into two steps. The first step involves optimizing the hash code matrix $\mathbf{B} \in \lbrace -1, 1 \rbrace ^{k \times n}$, which represents the hash codes for both the training set and the learning target in the subsequent step, where $n$ is the number of images of the training set. The second step focuses on learning the hash function.

\subsubsection{Optimization of the first step}
We define $\mathbf{V} \in \mathbb{R}^{k \times n}$  as the real-valued intermediate state of $\mathbf{B}$ and specify its optimization objective as follows:
\begin{equation} \label{op_b}
\begin{aligned}
&\mathop{\min}_{\mathbf{B,P,V,R}} \Vert \mathbf{Y} - \mathbf{PV} \Vert^{2}_{F} + \alpha \Vert \mathbf{B} - \mathbf{RV} \Vert^{2}_{F} \\
    &s.t ~\mathbf{B} \in \lbrace -1, 1 \rbrace^{k \times n}, ~\mathbf{R}^\top \mathbf{R} = \mathbf{I},
\end{aligned}
\end{equation}
here, $\Vert \cdot \Vert_{F}$ represents the Frobenius norm of a matrix, $\mathbf{Y} = \lbrace y_i \rbrace_{i=1}^{n} \in \lbrace 0, 1 \rbrace^{C \times n}$ represents the label matrix, where $C$ is the number of classes, $\mathbf{R} \in \mathbb{R}^{r \times r}$ is an orthogonal rotation matrix, and $\alpha$ is a hyperparameter.
Subsequently, the variables $\mathbf{P, V, R}$ and $\mathbf{B}$ can be optimized alternately as follows.

Optimize $\mathbf{P}$: By fixing all variables except for $\mathbf{P}$ and setting the derivative of Eq.~(\ref{op_b}) to zero, we can find that $\mathbf{P}$ has a closed-form solution as follows:
\begin{equation}
    \mathbf{P} = \mathbf{YV}^\top(\mathbf{YV}^\top)^{-1}.
\end{equation}

Optimize $\mathbf{V}$: Similarly, by fixing all variables except for $\mathbf{V}$ and setting the derivative of Eq.~(\ref{op_b}) to zero, we can find that $\mathbf{V}$  has a closed-form solution as follows:
\begin{equation}
     \mathbf{V} = (\mathbf{P}^\top\mathbf{P} +\alpha\mathbf{R}^\top\mathbf{R})^{-1} (\mathbf{P}^\top\mathbf{Y} +\alpha\mathbf{R}^\top\mathbf{B}).
\end{equation}

Optimize $\mathbf{R}$: When other variables except $\mathbf{R}$ are fixed, the Eq.~(\ref{op_b}) can be solved using Singular Value Decomposition~(SVD).
Given $\mathbf{BV}^\top = \mathbf{S}\Omega \widetilde{\mathbf{S}}^\top$, we find the solution as $\mathbf{R} = \mathbf{S}\widetilde{\mathbf{S}}^\top$.

Optimize $\mathbf{B}$: Keeping all variables except $\mathbf{B}$ constant, Eq.~(\ref{op_b}) can be rewritten as:
\begin{equation}
\begin{aligned}
&\mathop{\min}_{\mathbf{B}} \mathrm{Tr}(\mathbf{B}^\top(\mathbf{RV})) \\
   & s.t ~\mathbf{B} \in \lbrace -1, 1 \rbrace^{k \times n}.
\end{aligned}
\end{equation}
Consequently, $\mathbf{B}$ has a closed-form solution given by: $\mathbf{B} = \mathrm{sign}(\mathbf{RV})$.

The optimization process for matrix $\mathbf{B}$ continues according to the steps mentioned above until convergence or reaching a pre-defined number of iterations. The resulting optimized matrix 
$\mathbf{B}$ serves as the hash code for the training set and is utilized in the subsequent steps of hash function learning.

\subsubsection{Optimization of the second step}
The objective of the second step is to learn the hash function by optimizing the following:
\begin{equation}
    \mathcal{L}_{\mathrm{hash}} = \mathrm{MSE}(\hat{\mathbf{h}}, \mathbf{B}),
\end{equation}
where $\mathrm{MSE}(\cdot)$ represents the mean squared error loss function, and $\hat{\mathbf{h}}$ denotes the hash code generated during hash function training.

\subsection{Overall Training Objective}
To enhance EViT's sensitivity to subcategory-specific discrepancies, we employ a classification loss as auxiliary supervision:
\begin{equation}
    \mathcal{L}_{\mathrm{cls}} = \mathrm{CE}(\hat{y}, y),
\end{equation}
where $\hat{y}$ is the predicted label and $y$ is the ground-truth label of the input image $\mathbf{X}$. The overall loss function is then given by
\begin{equation} \label{loss}
    \mathcal{L} = \mathcal{L}_{\mathrm{hash}} + \beta\bigl(\mathcal{L}_{\mathrm{cls}} +\mathcal{L}_{\mathrm{drg}}\bigr) + \sigma\mathcal{L}_{\mathrm{dkt}},
\end{equation}
where $\beta$ and $\sigma$ are hyper-parameters balancing the relative contribution of each loss term.

\subsection{Out-of-Sample Extension}
After training, only the EViT is used to generate binary codes for previously unseen query images.
Given a query image $\mathbf{X}_q$, the class token $\mathbf{E}_{\mathrm{class}}^{q}$ is first extracted. Our EET then produces a binary hash code by
\begin{equation}
    \mathbf{b}_q = \mathrm{sign}\Bigl(\mathrm{FC}_{\mathrm{hash}}\bigl(\mathrm{FC}_{\mathrm{cls}}\bigl(\mathbf{E}_{\mathrm{class}}^{q}\bigr)\bigr)\Bigr),
\end{equation}
where $\mathrm{FC}_{\mathrm{hash}}$ and $\mathrm{FC}_{\mathrm{cls}}$ are fully connected layers constituting the classification and hash heads, respectively. By adopting this procedure, EET seamlessly extends to out-of-sample data while maintaining efficiency in large-scale fine-grained image retrieval.

\section{Experiments} \label{s5}
\begin{table*}
    \centering
    \caption{Comparisons with the state-of-the-art methods of mAP (\%) on three fine-grained datasets with code bits from 16 to 64. The best results are shown in boldface.}
    \begin{tabular}{cccccccccccccc}
    \toprule
          \multirow{2}{*}{Method} &\multirow{2}{*}{Backbone} & \multicolumn{4}{c}{CUB-200-2011}  & \multicolumn{4}{c}{Stanford Cars} &\multicolumn{4}{c}{NABirds} \\
    \cmidrule(lr){3-6} \cmidrule(lr){7-10} \cmidrule(lr){11-14}
         & &16bits & 32bits & 48bits & 64bits & 16bits & 32bits & 48bits & 64bits & 16bits & 32bits & 48bits & 64bits \\
         \midrule
    DPN~\cite{dpn} &ResNet18 &49.61 &69.86 &72.33 &73.95 &70.86 &78.76 &79.93 &81.22  &32.39 &51.70 &59.51 &62.97 \\
    CSQ~\cite{csq} &ResNet18 &54.41 &63.93 &67.61 &69.68 &51.31 &67.12 &72.97 &75.11  &21.20 &35.90 &38.27 &39.49 \\
    OrthoHash~\cite{oneloss}  &ResNet18 &54.38 &68.10 &70.25 &70.44 &67.94 &78.53 &79.64 &80.06  &48.55 &52.06 &55.57 &55.65\\
         % \midrule
    DsaH~\cite{dsah} &ResNet18 & 50.54 &54.74 &66.01 &67.02 &39.24 &60.58 &68.81 &70.27 &12.71 &15.29 &11.98 &19.94  \\
    DLTH~\cite{dlth} &ResNet18 & 61.93 &68.61 &70.16 &73.22 &71.75 &75.35 &78.59 &80.10 &35.30 &42.91 &47.47 &50.32  \\
    sRLH~\cite{srlh} &ResNet18 & 62.79 &69.94 &70.96 &71.93 &73.70 &82.70 &83.97 &84.04 &50.21 &61.41 &64.39 &65.74  \\
    FISH~\cite{fish} &ResNet18 &69.17 &73.13 &74.10 &74.13 &81.35 &84.07 &83.88 &83.92 &56.68 &62.39 &65.01 &65.06 \\
    SEMICON~\cite{semicon} &ResNet50  &48.86 &72.61 &79.67 &82.33  &33.14 &68.47 &77.69 &82.61 &13.74 &29.79 &42.94 &45.02 \\
    \textsc{DAHNet}~\cite{dahnet} &ResNet50 &71.53 &81.69 &83.98 &83.79  &68.90 &85.46 &88.49 &89.01 &29.47 &47.43 &56.46 &61.31 \\
    ViT-Small~\cite{deit} &ViT-Small &76.95 &81.63 &83.40 &83.84  &82.85 &87.56 &88.57 &89.33 &58.30 &69.13 &72.16 &73.50 \\
    MSViT~\cite{msvit} &MSViT-S &57.79  &55.60  &61.60  &63.66  &43.99  &44.67  &46.04  &55.47  &18.48  &24.17  &28.09  & 30.16 \\
    DVF~\cite{dvf} &ViT-Small &78.98   &82.10  &83.64  &85.01 &83.21  &87.21  &88.64  &89.17  &\textbf{63.75}  &\textbf{75.15}  &\textbf{77.14}  &\textbf{78.42}   \\
    \midrule
    EET &ViT-Small  &\textbf{79.85} &\textbf{83.00} &\textbf{84.02} &\textbf{85.05} &\textbf{83.58} &\textbf{87.77} &\textbf{89.43} & \textbf{89.68} &63.45 &73.91 &76.56 &77.60 \\
    \bottomrule
    \end{tabular}
    
    \label{tab:tabel2}
\end{table*}

\begin{table*}
    \centering
    \caption{Comparisons with the state-of-the-art methods of mAP (\%) on two large-scale fine-grained datasets with code bits from 12 to 48.}
    \begin{tabular}{cccccccccc}
    \toprule
          \multirow{2}{*}{Method} &\multirow{2}{*}{Backbone} & \multicolumn{4}{c}{Vegfru} & \multicolumn{4}{c}{Food101}  \\
    \cmidrule(lr){3-6} \cmidrule(lr){7-10} 
         & &12bits & 24bits & 32bits & 48bits & 12bits & 24bits & 32bits & 48bits \\
         \midrule
    ExchNet~\cite{exchnet} &ResNet50 &23.55 &35.93 &48.27 &69.30 &45.63&55.48&56.39&64.19 \\   
    A$^2$-Net~\cite{anet} &ResNet50 &25.52 &44.73 &52.75 &69.77 &46.44&66.87&74.27&82.13 \\
    SEMICON~\cite{semicon} &ResNet50 &30.32 &58.45 &69.92 &79.77 &50.00&76.57&80.19&82.44  \\
    A$^2$-Net$^{++}$~\cite{anet++} &ResNet50 &30.54 &60.56 &73.38 &82.80 &54.51&81.46&82.92&83.66  \\
    \textsc{DAHNet}~\cite{dahnet} &ResNet50 &56.11 &78.07 &82.19 &85.56 &67.67&79.38&82.05&83.11  \\
    FISH~\cite{fish} &ResNet50 &77.52 &83.75 &83.73 &83.28 &82.82&83.03&83.53&84.60  \\
    ViT-Small~\cite{deit} &ViT-Small &76.36  &83.83   &85.16  &86.54  &85.36 &87.94 &88.08 &88.55   \\
    MSViT~\cite{msvit} &MSViT-S &19.77  &27.95  &31.91  &37.15  &27.40  &37.10  &43.18  &48.45   \\
    DVF~\cite{dvf} &ViT-Small &80.48  &86.59   &87.12  &\textbf{88.57} &85.64  &\textbf{88.07}   &88.27  &\textbf{88.76}  \\
    \midrule
    EET &ViT-Small &\textbf{81.84} &\textbf{86.94 }&\textbf{87.77} &88.47 &\textbf{86.16} &88.01 &\textbf{88.31} &88.72 \\
    \bottomrule
    \end{tabular}
    
    \label{tab:tabel3}
\end{table*}

\begin{table}
    \centering
    \caption{Comparisons with the state-of-the-art methods of mAP (\%) on iNat2017 datasets with code bits from 16 to 64.}
    \begin{tabular}{cccccc}
    \toprule
          \multirow{2}{*}{Method} &\multirow{2}{*}{Backbone} & \multicolumn{4}{c}{iNat2017} \\
    \cmidrule(lr){3-6} 
         & &16bits & 32bits & 48bits & 64bits \\
         \midrule
    ViT-Small~\cite{deit} &ViT-Small &2.59  &4.54  &5.66  &6.28   \\
    DVF~\cite{dvf} &ViT-Small &2.77  &4.84  &5.75  &6.53  \\
    \midrule
    EET &ViT-Small &3.65  &5.69  &7.13  &8.08 \\
    \bottomrule
    \end{tabular}
    
    \label{tab:tabel4}
\end{table}

\subsection{Datasets}
We conduct experiments on six fine-grained benchmarks, including CUB200-2011~\cite{wah2011caltech}, Stanford Cars~\cite{car}, NABirds~\cite{HornBFHBIPB15}, VegFru~\cite{HouFW17}, Food101~\cite{BossardGG14}, and iNat2017~\cite{inat}.
\begin{enumerate}
\item The \textbf{CUB-200-2011} dataset consists of 11,788 images spread across 200 subcategories. The training set comprises 5,994 images, while the test set comprises 5,794 images.
\item The \textbf{Stanford Cars} dataset comprises 16,185 car images grouped into 196 subcategories. Its official training set contains 8,144 images, while the testing set contains 8,041 images.
\item The \textbf{NAbirds} dataset comprises 48,562 images spanning 555 subcategories. Its official train/test split assigns 23,929 images to the training set and 24,633 images to the test set.
\item The \textbf{VegFru} dataset comprises a substantial collection of 160,731 images, covering 200 vegetable subcategories and 92 fruit subcategories. Its official training set includes 29,200 images (100 images per subcategory), with an additional 14,600 images in the validation set and a substantial 116,931 images in the testing set.
\item The \textbf{Food101}  dataset comprises a significant collection of 101,000 images categorized into 101 food types. Its official train/test split allocates 750 images per subcategory for the training set and 250 images per subcategory for the testing set.
\item The \textbf{iNat2017} dataset consists of 675,170 images classified into 5,089 species categories. It is class-imbalanced, with an official train-validation split of 579,184 images for training and 95,986 images for validation.
\end{enumerate}

All evaluations follow their respective official train/test splits.

\subsection{Implementation Details and Evaluation Metrics} \label{detail}
In our experiments, we employ the ViT-small~\cite{vit} pre-trained on ImageNet1K as the backbone.
All input images are resized to $224 \times 224$. During the training stage, we use the SGD optimizer and implement cosine annealing as the optimization scheduler. We set the learning rate to 1e-2 for all datasets, except for iNat2017, where it is set to 5e-3. CUB-200-2011 and Stanford Cars are trained for 90 epochs, NABirds and VegFru for 60 epochs, Food101 for 30 epochs, and iNat2017 for 10 epochs, with a batch size of 64.
All experiments are conducted on a single RTX 3090 GPU.
The hyper-parameter $\gamma_j~(j = 1, 2, 3)$ in Section~\ref{mctc}, the hyper-parameter $\beta, \sigma$ in Eq.~(\ref{loss}) are set to $\frac{1}{2}, \frac{1}{2}, \frac{1}{4}$, 0.1 and 1.0 respectively.

We employ several common metrics to evaluate the performance of fine-grained image retrieval, including the mean Average Precision (mAP) and the Precision-Recall (PR) curve.

mAP denotes the average precision of the top $Q$ retrieved images. During the testing phase, $Q$ equals the size of the training set for all datasets.

The PR curve illustrates the overall retrieval performance. A larger area under the PR curve indicates better retrieval performance achieved by the method.

\begin{figure*}
    \centering
\includegraphics[width=\textwidth]{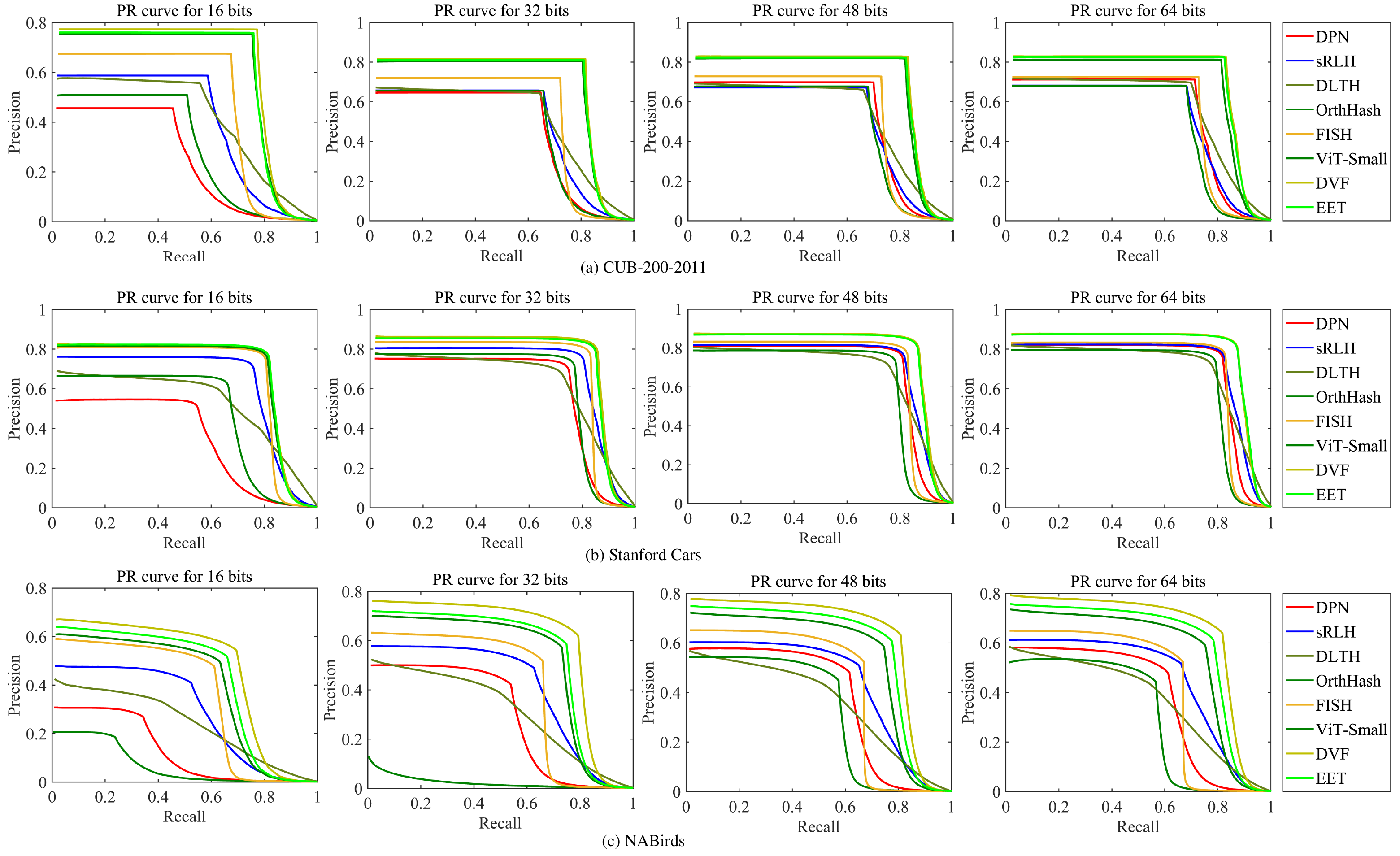}
    \caption{Precision-Recall curves of EET and state-of-the-art methods on the three datasets.}
    \label{fig:pr}
\end{figure*}

\begin{table*}
    \centering
    \caption{The mAP(\%) results and computational costs (GFlops) comparisons with methods based on ResNet-50~(SEMICON, \textsc{DAHNet}, FISH, CMBH), and methods based on ViT~(ViT-Small and DVF) with code bits from 16 to 64.}
    \begin{tabular}{lccccccccccccc}
    \toprule
    \multirow{2}{*}{Method} & \multicolumn{4}{c}{CUB-200-2011} & \multicolumn{4}{c}{VegFru} &\multicolumn{4}{c}{Food101} &\multirow{2}{*}{GFLOPs}\\
    \cmidrule(lr){2-5} \cmidrule(lr){6-9} \cmidrule(lr){10-13}
         & 12bits & 24bits & 32bits & 48bits  & 12bits & 24bits & 32bits & 48bits  & 12bits & 24bits & 32bits & 48bits \\
         \midrule
    SEMICON~\cite{semicon} &37.76 &65.41 &72.61 &79.67 &30.32 &58.45 &69.92 &79.77 &50.00&76.57&80.19&82.44 &7.09\\
    \textsc{DAHNet}~\cite{dahnet} &61.69 &79.00 &81.69 &83.98 &56.11 &78.07 &82.19 &85.56 &67.67&79.38&82.05&83.11  &10.59 \\
    FISH~\cite{fish} &71.90 &74.66 &75.44 &76.55 &77.52 &83.75 &83.73 &83.28 &82.82&83.03&83.53&84.60  &4.13  \\
    ViT-Small~\cite{deit} &72.41  &80.72   &81.93  &83.40  &76.36  &83.83   &85.16  &86.54 &85.36 &87.94 &88.08 &88.55 &4.25 \\
    DVF~\cite{dvf} &73.17  &79.94   &82.10  &83.64  &80.48  &86.59   &87.12  &88.57 &85.64  &88.07   &88.27  &88.76 &11.53  \\
    CMBH~\cite{cmbh} &84.07  &85.79  &86.21  &86.47  &84.37  &88.63   &88.46  &89.06 &87.85  &88.71   &89.28  &88.87 &4.34  \\
    EET & 75.02  & 82.68 &83.00 &84.02 &81.84&86.94 &87.77 &88.47 &86.16 &88.01 &88.31 &88.72 &3.07 \\
    \bottomrule
    \end{tabular}
    
    \label{tab:test_time}
\end{table*}

\subsection{Results and Analysis}
\subsubsection{Comparions Results}
We benchmark EET against several state-of-the-art deep hashing methods, namely DPN~\cite{dpn}, CSQ~\cite{csq}, OrthoHash~\cite{oneloss}, ViT-Small~\cite{deit}, MSViT~\cite{msvit} DSaH~\cite{dsah}, DLTH~\cite{dlth}, sRLH~\cite{srlh}, FISH~\cite{fish}, ExchNet~\cite{exchnet}, A$^2$-Net~\cite{anet}, SEMICON~\cite{semicon}, A$^2$-Net$^{++}$~\cite{anet++}, \textsc{DAHNet}~\cite{dahnet}, and DVF~\cite{dvf}.
Table~\ref{tab:tabel2} displays the mAP results for CUB-200-2011, Stanford Cars, and NABirds.
The mAP results for large-scale datasets VegFru, Food101, and iNat2017 are presented in Table~\ref{tab:tabel3} and Table~\ref{tab:tabel4}. 
The mAP metric shows that our method achieves competitive performance compared to other state-of-the-art methods on all datasets.

Overall, the performance of ViT-Small has achieved competitive or leading results compared to CNN-based methods across various datasets, demonstrating that the ViT architecture can indeed enhance performance. However, MSViT's performance on many datasets lags behind CNN-based methods, suggesting that the ViT architecture alone does not account for all performance improvements.
Meanwhile, the discriminative knowledge transfer and discriminative region guidance enhance the performance of EET, allowing it to surpass ViT-Small.
In addition, although DVF achieves comparable performance to ours, it introduces an additional grounding model to help object location, which incurs a lot of computational overhead. For details on computational costs, refer to Section~\ref{cost}.

\subsubsection{Comparison on Precision-Recall}
Additionally, we evaluate performance based on Precision-Recall (PR) across various hash bits and datasets. The experimental results are depicted in Figure~\ref{fig:pr}.
The proposed method achieves significant improvements over the baseline method ViT on the challenging NABirds dataset, demonstrating the superiority of EET.
Furthermore, EET achieves competitive performance compared with the state-of-the-art method DVF~\cite{dvf} on three datasets. The superior performance of DVF may be due to the larger proportion of objects in its input images, which makes it easier to focus on discriminative areas.

\subsubsection{Computational Costs Analysis} \label{cost}
As discussed in Section~\ref{s1}, hashing methods aim to strike a balance between retrieval accuracy and efficiency. Therefore, it is essential to consider computational costs during the inference phase, especially for large-scale image retrieval tasks. 
To assess the effectiveness and efficiency of our proposed method, we compare its retrieval performance and inference time with four state-of-the-art ResNet-50-based methods: SEMICON~\cite{semicon}, \textsc{DAHNet}~\cite{dahnet}, FISH~\cite{fish}, and CMBH~\cite{cmbh}, as well as three ViT-based methods: ViT-Small~\cite{deit}, MSViT~\cite{msvit}, and DVF~\cite{dvf}. The results are presented in Table~\ref{tab:test_time}. 
From Table~\ref{tab:test_time}, we observe that while our method slightly lags behind CMBH in retrieval performance, it remains competitive on larger datasets. This is primarily due to the fact that CMBH employs complex feature learning modules and enhancement strategies, which increase its computational load. In contrast, our method is simpler yet highly effective, offering a good trade-off between accuracy and efficiency.
Furthermore, our approach is significantly more computationally efficient than CMBH, especially in terms of GFLOPs, which is crucial for large-scale fine-grained image retrieval tasks. This efficiency makes our method particularly advantageous when scaling up to large datasets, where computational costs can become a limiting factor.

\begin{table*}
    \centering
    \caption{The mAP(\%) results and inference latency~(ms) of modules ablation study with code bits from 16 to 64.}
    \begin{tabular}{lccccccccccccc}
    \toprule
    \multirow{2}{*}{Configuration} & \multicolumn{4}{c}{CUB-200-2011} & \multicolumn{4}{c}{Stanford Cars} &\multicolumn{4}{c}{NABirds} &\multirow{2}{*}{Latency}\\
    \cmidrule(lr){2-5} \cmidrule(lr){6-9} \cmidrule(lr){10-13}
         & 16bits & 32bits & 48bits & 64bits & 16bits & 32bits & 48bits & 64bits & 16bits & 32bits & 48bits & 64bits \\
         \midrule
    Baseline &76.95 &81.93 &83.40 &83.84 &82.85 &87.56 &88.57 & 89.33 &58.30 &69.13 &72.16 &73.50 &0.82\\
    + CTP &76.10 &80.98 &82.70 &83.78 &79.96 &85.12 &86.66 & 87.86 &57.61 &68.68 &71.44 &72.51 &0.47\\
    + CTP + DKT &77.63 &81.88 &83.30 &84.05 &82.81 &86.77 &88.28 & 88.85 &60.10 &71.39 &74.67 &75.42 &0.47\\
    EET &\textbf{79.85} &\textbf{83.00} &\textbf{84.02} &\textbf{85.05} &\textbf{83.58} &\textbf{87.77} &\textbf{89.43} & \textbf{89.68} &\textbf{63.45} &\textbf{73.91} &\textbf{76.56} &\textbf{77.60} &0.47\\
    \bottomrule
    \end{tabular}
    \label{tab:ablation}
\end{table*}

\begin{table*}
    \centering
    \caption{Comparison with the raw attention of the Vision Transformer on three datasets with code bits from 16 to 64.}
    \begin{tabular}{lcccccccccccc}
    \toprule
    \multirow{2}{*}{Configuration} & \multicolumn{4}{c}{CUB-200-2011} & \multicolumn{4}{c}{Stanford Cars} &\multicolumn{4}{c}{NABirds} \\
    \cmidrule(lr){2-5} \cmidrule(lr){6-9} \cmidrule(lr){10-13}
         & 16bits & 32bits & 48bits & 64bits & 16bits & 32bits & 48bits & 64bits & 16bits & 32bits & 48bits & 64bits \\
        \midrule
    EET w/ Raw Attention &79.29	&82.67	&\textbf{84.35}	&84.74 &82.88	&87.09	&89.11	&89.24 &62.74	&73.13	&76.24	 &77.35 \\
    EET &\textbf{79.85} &\textbf{83.00} &84.02 &\textbf{85.05} &\textbf{83.58} &\textbf{87.77} &\textbf{89.43} & \textbf{89.68} &\textbf{63.45} &\textbf{73.91} &\textbf{76.56} &\textbf{77.60} \\
    \bottomrule
    \end{tabular}
    \label{tab:ablation_attention}
\end{table*}

\begin{table*}
    \centering
    \caption{The mAP(\%) results and inference latency~(ms) of different pruning positions with code bits from 16 to 64. $\lbrace 0 \rbrace$ denotes without pruning.}
    \begin{tabular}{lccccccccccccc}
    \toprule
    \multirow{2}{*}{Pruning Position} & \multicolumn{4}{c}{CUB-200-2011} & \multicolumn{4}{c}{Stanford Cars} &\multicolumn{4}{c}{NABirds} &\multirow{2}{*}{Latency}\\
    \cmidrule(lr){2-5} \cmidrule(lr){6-9} \cmidrule(lr){10-13}
         & 16bits & 32bits & 48bits & 64bits & 16bits & 32bits & 48bits & 64bits & 16bits & 32bits & 48bits & 64bits \\
         \midrule
    $\lbrace 0 \rbrace$ &80.18 &83.17 &84.52 &85.11 &85.24 &88.77 &89.71 & 90.30 &63.82 &74.10 &76.70 &77.85 &0.82\\
    $\lbrace 4, 8, 10 \rbrace$ &79.85 &83.00 &84.02 &85.05 &83.58 &87.77 &89.43 & 89.68 &63.45 &73.91 &76.56 &77.60 &0.47\\
     $\lbrace 4 \rbrace$ &79.96  &83.47  &84.50  &85.17  &84.27  &88.08  &89.42  &90.03   &63.58  &73.98  &76.62  &77.72 &0.54\\
     $\lbrace 8 \rbrace$ &79.81  &83.01  &84.31  &84.96  &84.50  &88.32  &89.40  &90.17   &63.29  &73.95  &76.59  &77.61 &0.61\\
     $\lbrace 10 \rbrace$ &79.80  &83.24  &84.36  &84.77  &85.09  &88.53  &89.92  &90.08   &63.78  &74.11  &76.71  &77.68  &0.69\\
     $\lbrace 4, 8 \rbrace$ &79.92  &83.01  &84.51  &84.75  &83.76  &87.97  &89.29  &89.98   &63.24  &73.85  &76.61  &77.49 &0.49\\
     $\lbrace 4, 10 \rbrace$ &79.96  &83.21  &84.66  &84.69  &83.58  &88.12  &89.50  &90.07   &63.57  &73.91  &76.61  &77.63 &0.50\\
     $\lbrace 8, 10 \rbrace$ &79.50  &83.19  &84.35  &84.75  &84.93  &88.07  &89.59  &89.92   &63.40  &73.83  &76.71  &77.53 &0.60\\
    \bottomrule
    \end{tabular}
    \label{tab:token_position}
\end{table*}

\subsubsection{Ablatuion Studies}
To validate the effectiveness of each component in our proposed framework, we decompose EET into its constituent modules and perform ablation experiments on the CUB-200-2011, Stanford Cars, and NABirds datasets. Specifically, EET comprises three key components: Content-based Token Pruning (CTP) (Section~\ref{mctc}), Discriminative Knowledge Transfer (DKT) (Section~\ref{dkt}), and Discriminative Region Guidance (DRG) (Section~\ref{drg}). Additionally, we also include a \textbf{Baseline} model that excludes all three modules.

From the results in Table~\ref{tab:ablation}, we make the following observations:
\begin{itemize}
    \item \textit{Efficiency from CTP:}~CTP substantially improves the model’s inference efficiency by pruning tokens. However, it also causes a notable performance drop, especially on the Stanford Cars dataset. This may be because token pruning removes critical features for car images, which often rely on subtle distinctions (\emph{e.g.,}~car fronts, headlights, or logos).
    \item \textit{Discriminative Power from DKT and DRG:}~DKT and DRG significantly enhance the model’s ability to recognize fine-grained objects, enabling EET to outperform the standard ViT on multiple datasets. In particular, these modules yield considerable gains on the challenging NABirds dataset, underscoring their capacity to capture subtle inter-class differences. Moreover, since neither DKT nor DRG increases inference time, EET retains its efficiency advantage despite the added discriminative power.
\end{itemize}
Overall, these ablation results illustrate the complementary roles of CTP, DKT, and DRG in achieving a strong trade-off between accuracy and efficiency in large-scale fine-grained image retrieval.

\subsubsection{Comparison of Attention Mechanism in ViT}
In this part, we conduct comparative experiments to evaluate how different token-importance calculation methods in CTP influence performance. Specifically, we compare raw attention in vanilla ViT against attention weighted by token content importance in our EET. As shown in Table~\ref{tab:ablation_attention}, adopting a content-weighted strategy yields notable performance gains, indicating that placing greater emphasis on informative tokens can significantly enhance retrieval accuracy. Moreover, the additional computational overhead introduced by weighting token attention is negligible, further validating the practicality of our method.

\subsubsection{The Effect of Hierarchical Token Pruning}
We conduct experiments with various pruning combinations to verify the efficiency gains brought by hierarchical insertion of CTP, and demonstrate the current strategy is effectiveness-efficiency optimal. The results are given in Table~\ref{tab:token_position}. The pruning strategy, with token pruning positions at $\lbrace$4, 8, 10$\rbrace$, shows significant efficiency advantages while maintaining effectiveness compared to without pruning. This clearly demonstrates the effectiveness and efficiency of our method. Additionally, its performance is comparable to the most effective pruning strategy with a token pruning position at $\lbrace$10$\rbrace$. This is because we preserve discriminative tokens by pruning at a lower rate (0.5) in the shallower layers 4 and 8. This confirms the scarcity of discriminative tokens crucial for FGIR. Moreover, in the Stanford Cars dataset, significant performance improvement is observed upon removing the token pruning from the 4th layer. This may be attributed to the regular shape of cars, where pruning tokens from shallow layers may lead to a loss of discriminative information, thereby affecting retrieval accuracy.

\subsubsection{Pruning Ratio of Each Pruning Position}
The pruning ratio of our CTP at each pruning position needs to be considered. We experiment with several pruning ratio configurations to demonstrate that the current strategy optimizes performance and inference time. The results are given in Table~\ref{tab:token_ration}. As the compression ratio rises, the efficiency of EET also increases gradually; however, reducing it to $\frac{1}{64} (\frac{1}{4} \times \frac{1}{4} \times \frac{1}{4})$ leads to a notable performance decline. This may be attributed to excessive compression, leading to the loss of too many informative tokens and consequently impacting retrieval accuracy. Considering both effectiveness and efficiency, we set $N_1$, $N_2$ and $N_3$ to $\frac{1}{2}$, $\frac{1}{2}$ and $\frac{1}{4}$ respectively for all datasets.

\begin{table*}
    \centering
    \caption{The mAP(\%) results and inference latency~(ms) of different pruning ration with code bits from 16 to 64.}
    \begin{tabular}{cccccccccccccc}
    \toprule
    \multirow{2}{*}{Ratio~($N_{1}, N_{2}, N_{3}$)} & \multicolumn{4}{c}{CUB-200-2011} & \multicolumn{4}{c}{Stanford Cars} &\multicolumn{4}{c}{NABirds} &\multirow{2}{*}{Latency}\\
    \cmidrule(lr){2-5} \cmidrule(lr){6-9} \cmidrule(lr){10-13}
         & 16bits & 32bits & 48bits & 64bits & 16bits & 32bits & 48bits & 64bits & 16bits & 32bits & 48bits & 64bits \\
        \midrule
    $\lbrace 1, 1, 1 \rbrace$ &80.18 &83.17 &84.52 &85.11 &85.24 &88.77 &89.71 & 90.30 &63.82 &74.10 &76.70 &77.85 &0.82\\
    $\lbrace \frac{1}{2}, \frac{1}{2}, \frac{1}{2} \rbrace$ &79.91  &82.86  &84.28  &85.22  &83.49  &87.59  &89.39  &89.75   &63.47  &73.89  &76.55  &77.59 &0.55\\
    $\lbrace \frac{1}{2}, \frac{1}{2}, \frac{1}{4} \rbrace$ &79.85 &83.00 &84.02 &85.05 &83.58 &87.77 &89.43 & 89.68 &63.45 &73.91 &76.56 &77.60 &0.47\\
    $\lbrace \frac{1}{2}, \frac{1}{4}, \frac{1}{4} \rbrace$ &78.83  &82.28  &83.76  &83.96  &83.14  &87.51  &88.82  &89.36   &62.98  &73.40  &76.11  &77.14 &0.45\\
    $\lbrace \frac{1}{4}, \frac{1}{4}, \frac{1}{4} \rbrace$ &76.91  &81.44  &82.53  &83.38  &80.20  &85.26  &87.01  &87.61   &61.57  &72.22  &75.04  &75.88  &0.38\\
    \bottomrule
    \end{tabular}
    \label{tab:token_ration}
\end{table*}

\begin{figure}[t!]
    \centering
    \includegraphics[width=0.75\linewidth]{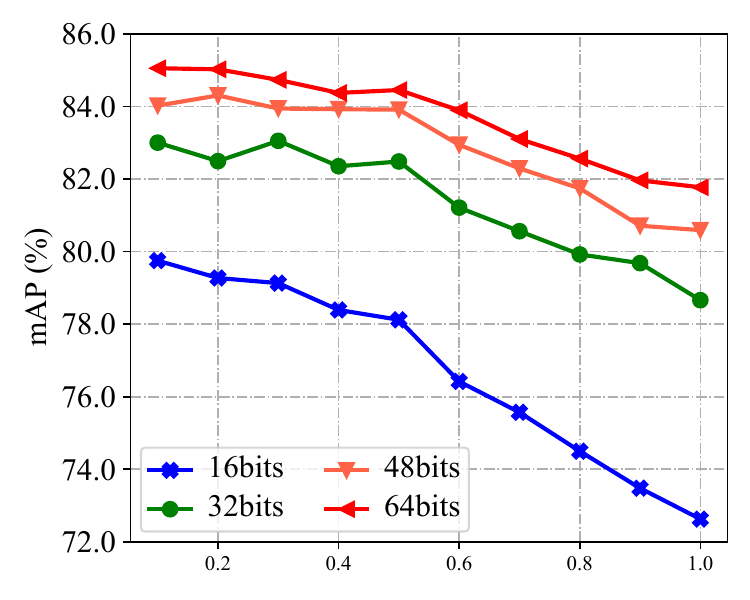}\vspace{-2mm}
    \caption{The hyper-parameter analysis of $\beta$ on CUB-200-2011 with code bits from 16 to 64.}
    \label{fig:beta}
\end{figure}

\begin{figure}[t!]
    \centering
    \includegraphics[width=0.75\linewidth]{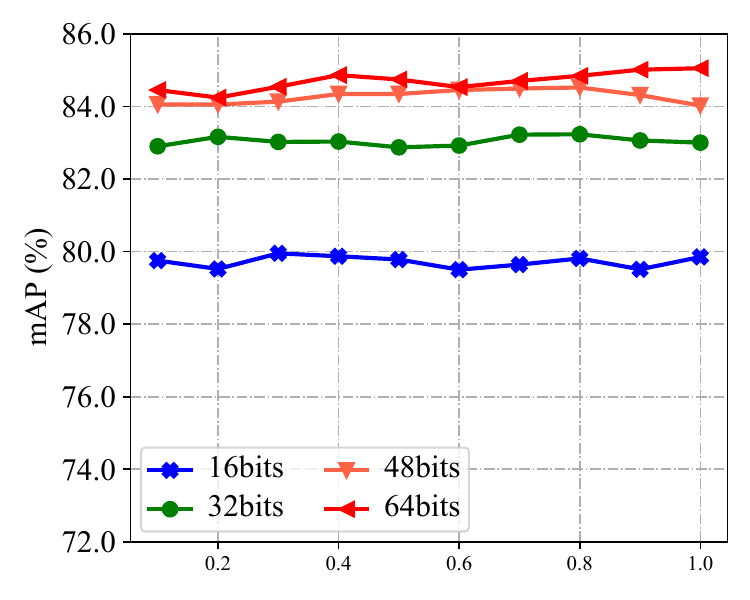}\vspace{-2mm}
    \caption{The hyper-parameter analysis of $\sigma$ on CUB-200-2011 with code bits from 16 to 64.}
    \label{fig:gamma}
\end{figure}

\subsubsection{Hyper-parameter $\beta$ Analysis}

The parameter $\beta$ in our method determines the proportion of feature representation learning. In this section, we analyze the impact of $\beta$ on the mAP for the CUB-200-2011 dataset.
Figure~\ref{fig:beta} shows the mAP results with different $\beta$ values, with other settings kept constant. The figure indicates that performance degrades as $\beta$ increases, with the worst results at $\beta$ = 1.0. This occurs because a high $\beta$ value hinders the overall optimization of the model, particularly the hash code learning. Therefore, we set the hyper-parameters $\beta$ = 0.1 in Eq.~(\ref{loss}) for all datasets. 

\begin{figure*}[t!]
    \centering
    \includegraphics[width=\textwidth]{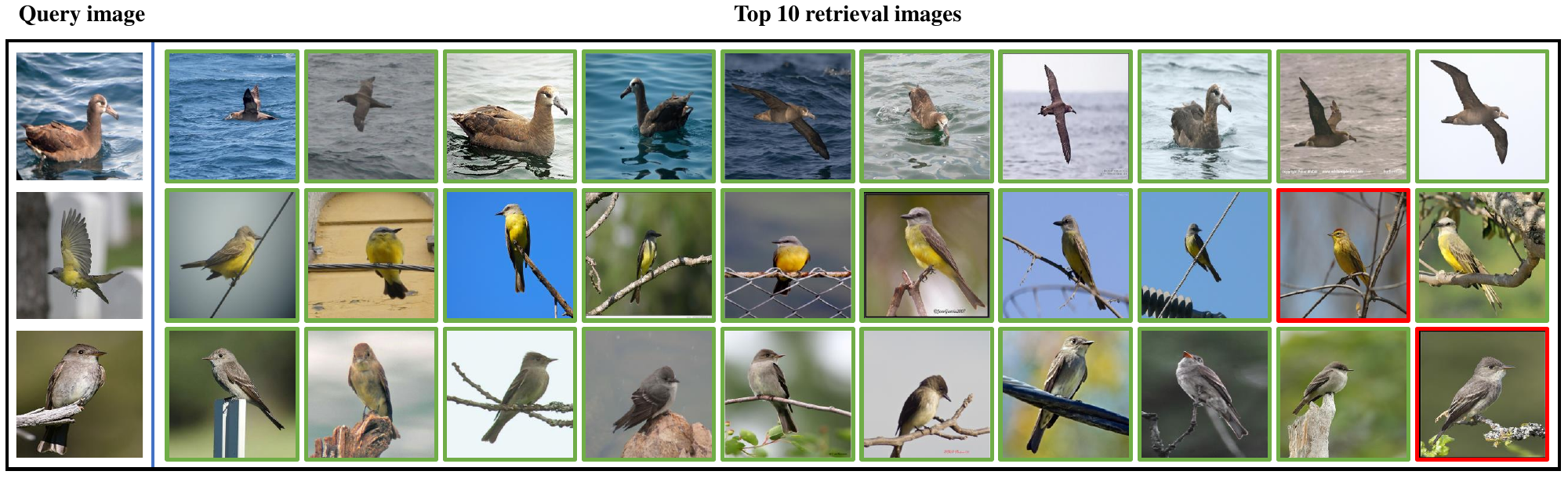}
    \caption{Examples of top 10 retrieval samples of the proposed EET on the CUB-200-2011. The retrieval images with green boxes are the correct ones, and those with red boxes are the wrong ones.}
    \label{fig:retri}
\end{figure*}

\begin{figure}[t!]
    \centering
    \includegraphics[width=0.9\linewidth]{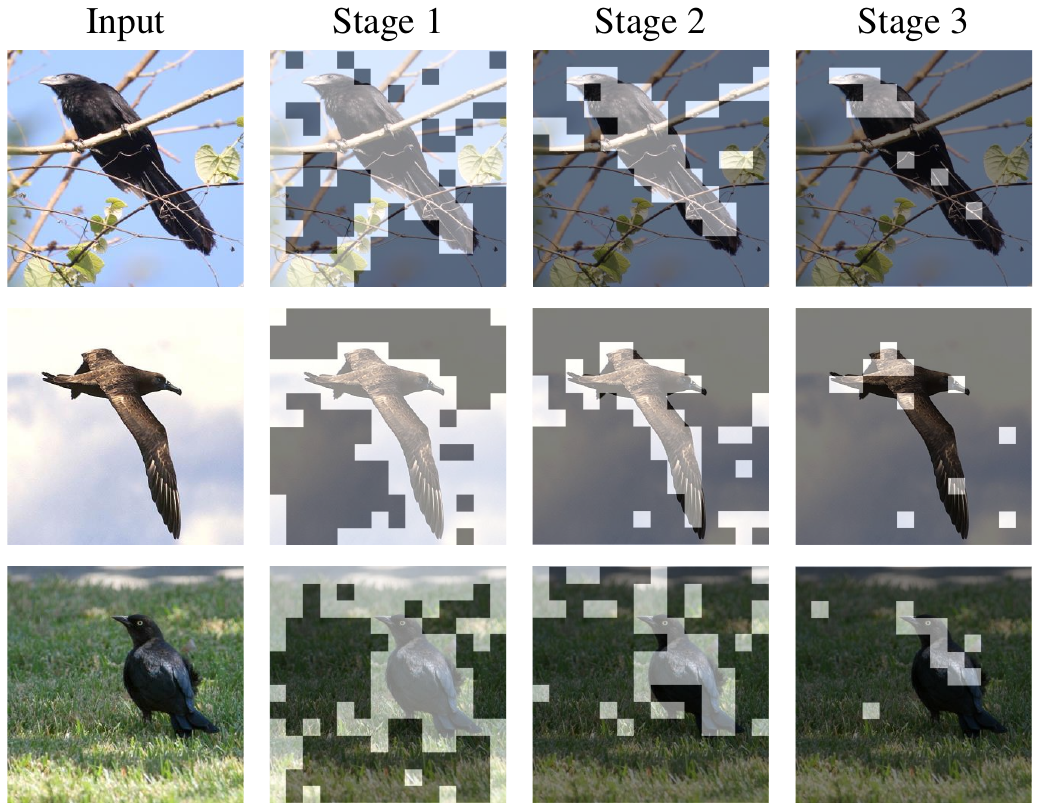}
    \caption{Visualization results of token pruning for samples from CUB-200-2011.}
    \label{fig:visual}
\end{figure}

\begin{figure}[t!]
    \centering
    \includegraphics[width=0.95\linewidth]{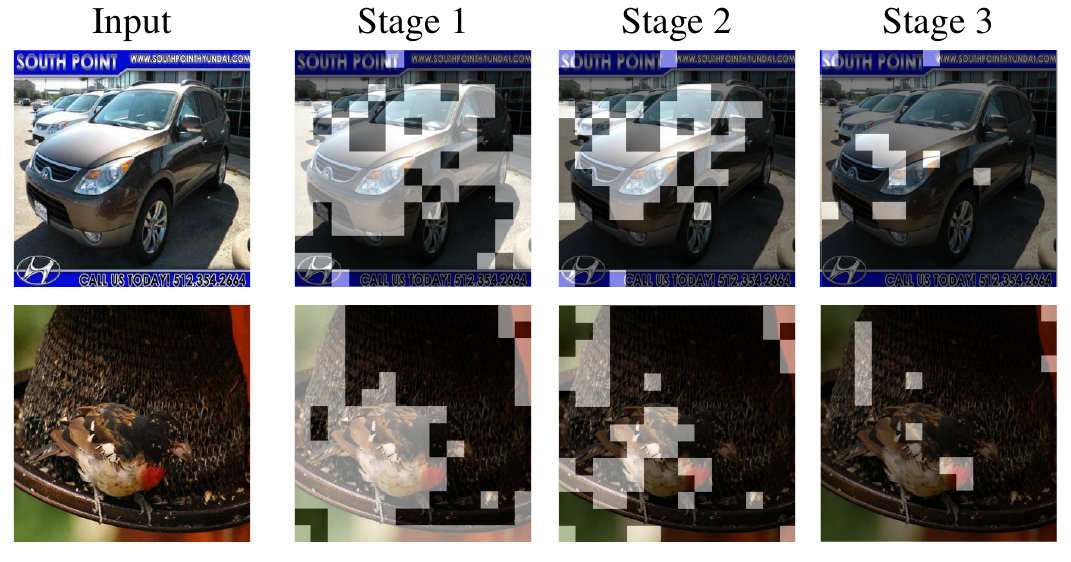}
    \caption{Failure cases of the CTP on Stanford Cars and CUB-200-2011 datasets.}
    \label{fig:fail}
\end{figure}

\subsubsection{Hyper-parameter $\sigma$ Analysis}
To verify the impact of the hyperparameter $\sigma$ on the experimental results, we conducted an ablation experiment.  When we analyze the influence of $\sigma$, the remaining parameters are set and as the default values in Section~\ref{detail}. The results are shown in Figure~\ref{fig:gamma}. The figure shows that when $\sigma$ is greater than 0, it leads to performance gains, and the size of $\sigma$ has little impact on the results. This demonstrates the robustness of the proposed discriminative knowledge transfer module. Therefore, we set the hyperparameter $\sigma$ to 1.0 in Eq.~(\ref{loss}) for all datasets.

\subsubsection{Visualization}
To further understand the interpretability of EET, we conducted a visualization analysis of the intermediate process of our token pruning, as shown in Figure~\ref{fig:visual}. In the early stage, EET mainly discards meaningless tokens such as background tokens, and in the later stage, it discards low-discriminative tokens in the object region. This aligns with our expectations: background tokens, receiving lower attention, are discarded early, while low-discriminative tokens within the object are gradually discarded later.

\subsection{Limitations}
While the proposed EET framework demonstrates significant improvements in FGIR, certain limitations remain, particularly in challenging scenarios where subtle distinctions between images may lead to errors in retrieval.

As shown in Figure~\ref{fig:retri}, we visualize the top 10 retrieval results of our method on the CUB-200-2011 dataset. EET retrieves accurate results even when there is a considerable difference between the query image and the retrieved images. However, some retrieval errors are visible, and in these cases, even human observers may struggle to distinguish between the correct and incorrect results. These errors are indicative of the challenges involved in fine-grained retrieval tasks, where certain subtle features may not be captured effectively. To address this, further advancements in the model's discriminative power are necessary, possibly through more advanced attention mechanisms or hybrid architectures that can better highlight critical fine-grained features.

Figure~\ref{fig:fail} illustrates several failure cases arising from the content-based token pruning (CTP) module of EET. In the first row, we observe that CTP fails to focus on key regions such as the car’s front, headlights, and logo, which are essential for accurate retrieval in the Stanford Cars dataset. This issue may explain the performance drop observed when combining CTP with the baseline model on this dataset. In the second row, CTP is influenced by background elements that resemble the object of interest, causing the model to mistakenly focus on the background rather than the object itself. This is a known challenge in image retrieval, where background clutter can mislead the attention mechanism.

Despite these limitations, EET's ability to balance efficiency and effectiveness makes it particularly advantageous for large-scale fine-grained image retrieval tasks. In the future, these failure cases will guide the refinement of the model, improving its robustness and reliability.
\section{Conclusion} \label{s6}
This paper addresses the challenges of applying Vision Transformers to large-scale fine-grained image retrieval and introduces an efficient and effective Vision Transformer framework, EET. Our model is specifically designed to be both efficient and effective. Specifically, by hierarchically incorporating content-based token pruning into ViT, we create EViT, which enables efficient processing of large-scale, fine-grained data. Additionally, the discriminative transfer strategy, comprising both discriminative knowledge transfer and discriminative region guidance, allows EViT to effectively distinguish fine-grained objects within subcategories, without adding extra computational cost.
We conduct comparative experiments and comprehensive ablation studies on multiple fine-grained image retrieval datasets, demonstrating the superior efficiency and effectiveness of EET. In the future, we aim to extend this framework to other Vision Transformer architectures, such as the Swin Transformer, and apply it to more challenging tasks, including fine-grained sketch-based image retrieval and unsupervised fine-grained retrieval.

\bibliography{ref}
\bibliographystyle{IEEEtran}

\end{document}